\documentclass[preprint,11pt]{elsarticle}     
\usepackage{amsmath} 
\usepackage{bm}
\usepackage{float}  
\usepackage{lineno}     
\usepackage[colorlinks=true,linkcolor=blue,allcolors=blue]{hyperref}   
\usepackage[margin=2.5cm]{geometry}

\usepackage{amssymb}

\newcommand{\Hbulk}{ H_{\mathrm{bulk}} }

\newcommand{\Hmf}{ H_{\textrm{MF}} }              
\newcommand{\Zmf}{ Z_{\textrm{MF}} }     
\newcommand{\Fmf}{ F_{\textrm{MF}} }

\newcommand{\Nb}{ N }    
\newcommand{\Ntherm}{ N_{\mathrm{therm}} } 
\newcommand{\NMC}{ N_{\mathrm{MC}} } 
\newcommand{\NIC}{ N_{\mathrm{IC}} } 
\newcommand{\prodB}{ \prod_{i}^{\Nb} } 
\newcommand{\sumB}{ \sum_{i}^{ \Nb } }      
\newcommand{\sumStates}{ \sum_{\sigma_i=\pm1}^{} }     
\newcommand{\hsrd}{ h_{s}^{\prime} }     
\newcommand{\Jrd}{ J^{\prime} }       

\usepackage{xcolor}   

\newcommand{\mycomment}[1]{}    

\mycomment{ 
\usepackage{caption}  
\DeclareCaptionFont{myfont}{\fontsize{10pt}{10pt}\selectfont}  
\captionsetup{
  font=myfont,
  labelfont=bf,
  justification=justified,
  format=plain
}  
}  
 
\begin{document}

\begin{frontmatter}



\title{ Modeling public opinion control by a charismatic leader }

       
\author[inst1]{Tiago S. A. N. Sim\~{o}es} 

\affiliation[inst1]{organization={University of Campania “Luigi Vanvitelli”, Department of Mathematics and Physics},
            addressline={Caserta, Viale Lincoln, 5, 81100},  
            country={Italy}}   

\author[inst2]{Antonio Coniglio} 
\affiliation[inst2]{organization={Università degli Studi di Napoli ‘‘Federico II’’, Physics Department},
            addressline={I-80126 Napoli}, 
            country={Italy}} 

\author[inst3,inst4]{Hans J. Herrmann}
             
\affiliation[inst3]{organization={Universidade Federal do Ceará, Departamento de Física},
            addressline={Fortaleza, Ceará, 60451-970},  
            country={Brazil}}   
\affiliation[inst4]{organization={ESPCI, PMMH},
            addressline={Paris, 7 quai St. Bernard, 75005 },   
            country={France}}     

\author[inst1]{Lucilla de Arcangelis}

\begin{abstract}
We study the average long-time behavior of the binary opinions of a social group with peer-to-peer interactions under the influence of an external bias and a persuadable leader, a strongly-biased agent with a dynamic opinion with the intention of spreading it across the system. We use a generalized, fully-connected Ising model, with each spin representing the binary opinion of an agent at a given time and a single, super spin representing the opinion of the leader. External fields and interaction constants model the opinion bias and peer-to-peer interactions, respectively, while the temperature $T$ models an idealized social climate, representing an authoritarian regime if $T$ is low or a liberal one if $T$ is high.  
We derive a mean-field solution for the average magnetization $m$, the “social mood”, and investigate how $m$ and the super spin magnetization vary as a function of $T$.  
We find that, depending on the initial conditions, due to the presence of metastable states, the sign of the average magnetization depends on the temperature. Finally, we verify that this effect is also present even if we consider only nearest-neighbor interactions within the social group. 
\end{abstract}
 
\mycomment{
\begin{graphicalabstract}
\includegraphics{grabs}
\end{graphicalabstract} 

\begin{highlights}
\item Research highlight 1
\item Research highlight 2
\end{highlights} 
} 


\end{frontmatter}


\section{Introduction} 
\label{sec:Introduction} 

Social interactions between humans are an example of a complex multi-component system. While the behavior of a single individual is quite complex, the collective behavior emerging from the interactions among peers might produce global patterns, which are independent of the characteristics of each individual \cite{galam_sociophysics_1982,jones_simulation_1985,castellano_statistical_2009}. The Ising model has found application in social science, since it was first purposed by Weidlich \cite{weidlich_statistical_1971} 50 years ago to analyze the statistics of opinion polarization. Later, the model was also applied to study critical properties in the collective behavior in a strike \cite{galam_sociophysics_1982}, and, more recently, a generalized version of this model considering clustered networks was used to study the evolution of binary opinions in society \cite{baldassarri_ising_2023}. A number of other statistical models have been developed to analyze collective social phenomena, which include the models proposed by Schelling \cite{schelling_dynamic_1971} and Axelrod \cite{axelrod_dissemination_1997}, and more physical ones, such as the voter model \cite{clifford_model_1973,holley_ergodic_1975,redner_reality-inspired_2019}, the Sznajd model \cite{sznajd-weron_opinion_2000,sznajd-weron_sznajd_2005} and the majority-voter model \cite{de_oliveira_isotropic_1992}, which has been shown to have the same critical exponents as the Ising model \cite{vilela_majority-vote_2019}.  
Within the conceptual framework of social dynamics, a spin $i$ represents an individual or agent, with its state or spontaneous magnetization $ \sigma_i $ being its current (binary) opinion. An external field $h_i$ quantifies its bias to a certain opinion, while the interaction constants $J_{ij}$ quantify the peer-to-peer conditioning on personal opinions \cite{baldassarri_ising_2023}.  The temperature $ T $ can be interpreted as a "climate parameter" \cite{weidlich_statistical_1971}, quantifying the random sources of external conditioning that may influence an individual opinion. In a simplified view, large $ T $ could be interpreted as a "liberal" regime, where opinions tend to fluctuate, while low $ T $ corresponds to a "totalitarian" regime, where opinion changes are more difficult to achieve \cite{weidlich_statistical_1971}. The Ising model can easily be modified to account for extra features. For instance, heterogeneity can be introduced in the form of a so-called "zealot" \cite{mobilia_does_2003} or, sometimes, "stubborn" \cite{yildiz_discrete_2011} agent: an agent with a fixed opinion \cite{verma_impact_2014,khalil_zealots_2018} or with an opinion that evolves independently from its peers \cite{mobilia_does_2003}, with the intention of spreading it across the system. It is our purpose to investigate the effect of such a charismatic leader on the evolution of the global opinion as function of the degree of liberalism and initial conditions.   
  
In this work, we study a generalized ferromagnetic Ising model which represents an idealized social group biased towards some specific opinion with peer-to-peer interactions among all possible spin pairs, as well as a super spin, a zealot-like agent with a large but finite bias to a contrarian opinion interacting with all other spins. A possible consequence of the introduction of zealots is the appearance of steady states where consensus (all spins in the same state) is not achieved \cite{mobilia_role_2007,klamser_zealotry_2017}, enriching the phenomenology seen in this kind of models.  
A crucial novelty in our model is that the super spin, unlike traditional zealots spins, also has the possibility to change its state or opinion over time due to the interactions with the social group.    
We will focus on the equilibrium behavior of the "social mood" \cite{sznajd-weron_opinion_2000}, i.e. the average magnetization of the Ising model, as well as the particular average magnetization of the super spin, as a function of the climate parameter $T$ and the initial distribution of opinions. 
The presence of heterogeneity in the model might give rise to metastable states, in which the system might reside for long times, giving rise to opinion distributions that do not minimize the energy of the system but are nonetheless important to consider due to their persistence \cite{baldassarri_ising_2023}.   
We also study the behavior of the model in systems with finite size, as it is usually the case in opinion models \cite{toral_finite_2007,castellano_statistical_2009,khalil_zealots_2018}. This is because, in the context of social science, where the basic elements are humans, the system sizes considered are much smaller (at most of the order $ \sim 10^{9}$ if we consider the current global scale \cite{ritchie_population_2023}) when compared to the ones considered in material science, where the elements are instead atoms or molecules, being of the order of the Avogadro constant $ \sim 10^{23} $. Finally, we also consider a case where interactions in the social group are short-ranged.      
 
The rest of the manuscript is organized as follows: section \ref{sec:MFsolution} describes the model in more detail and presents a mean field solution for the average magnetization. In section \ref{sec:Numerical-results} we study finite-sized systems using Monte Carlo simulations and compare numerical solutions with the mean field solution. Section \ref{sec:Conclusions} then summarizes the main results and presents possible future directions for the study of this model. 

\section{Model}  
\label{sec:MFsolution} 
 
We consider a system made of $N$ bulk spins, under the influence of a weak external, negative field $h < 0$, with ferromagnetic connections $J > 0$ between every pair of bulk spins. An additional super spin $s$, under the influence of a local, strong positive field $ h_s \gg |h|$ is connected to all bulk spins through a strong ferromagnetic connection $J_s \gg J $.   
The Hamiltonian $H$ of this generalized Ising model can be decomposed into two main components,   
\begin{equation} \label{eq:H} 
    	H = H_s + \Hbulk  \textrm{ ,} 
\end{equation}   
with $H_s$ and $\Hbulk$ referring to the super spin $s$ and the bulk spins, respectively, 
\begin{align}  
\label{eq:Hs}    H_s &= - h_s \sigma_s - J_s \sigma_s \sum_{i}^{\Nb} \sigma_i  \textrm{ ,}  \\ 
\label{eq:Hbulk}	\Hbulk &= - h \sum_{i}^{\Nb} \sigma_i - \frac{J}{2} \sum_{ i \neq j }^{ N } \sigma_i \sigma_j  \textrm{ ,}      
\end{align} 
where $\sigma_i \in [-1,1]$ is the state of bulk spin $i$, $\sigma_s \in [-1,1]$ is the state of the super spin $s$ and the $1/2$ in the last term of (\ref{eq:Hbulk}) takes into account the double counting of the pairs $ij$.
The probability $P( \bm{ \sigma } )$ to observe a particular state $ \bm{ \sigma } = \{ \sigma_1, \sigma_2, ..., \sigma_N, \sigma_s \} $ at equilibrium is given by the Boltzmann distribution,  
\begin{equation}  
    P( \bm{ \sigma } ) = \frac{ e^{ - \beta H } }{ Z } \textrm{ ,}     
\end{equation} 
where $ \beta = 1 / (k_B T) $, with $ k_B $ being the Boltzmann constant and $ T $ the temperature, and $Z = \sum_{ \bm{\sigma} } e^{ - \beta H }$ is the partition function, where $\sum_{ \bm{\sigma} }$ denotes a sum over all possible states $\bm{ \sigma }$.    
                             
Within the conceptual framework of social dynamics, the state $ \sigma_i $ represents the current (binary) opinion of agent $i$. The bulk is the social group, with the external field $h < 0$ quantifying the intrinsic tendency or the exposure of this group towards a specific opinion \cite{baldassarri_ising_2023}, with the negative sign being just a convention, while $J$ models the peer-to-peer interactions between bulk spins \cite{baldassarri_ising_2023}, with a tendency for consensus since $J > 0$. 
The super spin, on the other hand, represents a leader with a strong ($h_s \gg |h|$) and contrarian ideology ($h_s$ with opposite sign to $h$) to the rest of the social group, and has the means to be persuasive or indoctrinate efficiently ($J_s \gg J$), seeking to align the opinion of its peers with its own ($J_s > 0$).      
         
 
To solve this generalized Ising model, we will use the mean field approximation and assume that the sum of all the fluctuations around the average magnetization are negligible,  
\begin{equation} \label{eq:mfApprox}  
    \sum_{ i \neq j }^{N} ( \sigma_i - m ) \cdot ( \sigma_j - m ) \approx 0 \textrm{ ,}     
\end{equation} 
where $ m \equiv \langle \sigma_i \rangle $, $\forall i$, is the average magnetization per spin. 
To proceed, we can start by adding and subtracting the average magnetization $m$ in the last term of Eq. (\ref{eq:Hbulk}),  
\begin{align*}
    \sum_{ i \neq j }^{ N } \sigma_i \sigma_j 
    & = \sum_{ i \neq j }^{ N } ( (\sigma_i - m) + m ) \cdot ( (\sigma_j - m) + m ) = \\ 
\refstepcounter{equation}\tag{\theequation}    & =  - N \cdot (N - 1) m^2 + 2m \cdot (N-1) \sum_{i}^{N} \sigma_i  
 + \sum_{ i \neq j }^{N} ( \sigma_i - m ) \cdot ( \sigma_j - m ) \textrm{ ,} \label{eq:pairTerm}       
\end{align*} 
where in the last step we expanded the product and wrote explicitly the results of the sums $\sum_{ i \neq j }^{ N } \sigma_i = \sum_{ i \neq j }^{ N } \sigma_j = (N-1) \sum_{i}^{N} \sigma_i$ and $\sum_{ i \neq j }^{ N } m^2 = N \cdot (N - 1) m^2$. 
Using approximation (\ref{eq:mfApprox}), we can disregard the last term in (\ref{eq:pairTerm}), and replace the result into Eq. (\ref{eq:Hbulk}), approximating the Hamiltonian (\ref{eq:H}) as     
\begin{equation} \label{eq:Hmf} 
   H \approx \Hmf = h_b - h_s \sigma_s - \left( J_s \sigma_s + h + J m \cdot ( \Nb - 1 ) \right) \sum_{i}^{ \Nb } \sigma_i \textrm{ ,}        
\end{equation}   
where $ h_b \equiv J m^{2} \Nb \cdot (\Nb - 1) / 2 $ is a term that sets the energy scale.  
Notice that, in the thermodynamic limit $N \rightarrow \infty$, the last term of Eq. (\ref{eq:pairTerm}) is effectively zero when compared to the first two, which are both proportional to $N$. Therefore the mean field Hamiltonian (\ref{eq:Hmf}) becomes exact when $N \rightarrow \infty$.     
  
We can then define a "mean-field" partition function $\Zmf$ by calculating $\Zmf = \sum_{ \bm{\sigma} } e^{- \beta \Hmf } $, where $ \sum_{ \bm{\sigma} } = \prod_{i}^{\Nb} \left( \sum_{\sigma_{i}=\pm1}^{} \right) \sum_{\sigma_{s}=\pm1}^{} = \sum_{\sigma_1=\pm1}^{} \sum_{\sigma_2=\pm1}^{} ... \sum_{\sigma_N=\pm1}^{} \sum_{\sigma_{s}=\pm1}^{} $. To proceed it is useful to expand the summation over the two super spin $\sigma_s = \pm 1$ states. We can then obtain the following expression for $\Zmf$ (see \ref{section:ZmfDerivation}),       
\begin{equation} \label{eq:Zmf} 
    \Zmf = e^{- \beta h_b } \cdot \left( 
    \left[ e^{\beta A} + e^{\beta B} \right]^{\Nb}   
    + \left[ e^{\beta C} + e^{\beta D} \right]^{\Nb} \right) \textrm{ ,}     
\end{equation} 
with 
\begin{align}   
    A &\equiv \hsrd + h_{+} \textrm{ ,}  \\   
    B &\equiv \hsrd - h_{+} \textrm{ ,}  \\    
    C &\equiv - \hsrd + h_{-} \textrm{ ,} \\    
    D &\equiv - \hsrd - h_{-} \textrm{ ,}  
\end{align}  
where $h_{+} = h + J_s + \Jrd m$ ($ h_{-} = h - J_s + \Jrd m$) is the effective field felt by the bulk spins when the super spin is up (down), and we introduced the definitions $ \hsrd \equiv h_s / \Nb $ and $ \Jrd \equiv (\Nb-1) J $.     
From (\ref{eq:Zmf}) we can define the probability $p(+)$ or $p(-)$ for the super spin to be up or down, respectively  
\begin{align}
    p(+) &= e^{- \beta h_b } \left[ e^{\beta A} + e^{\beta B} \right]^{\Nb} / \Zmf \textrm{ ,}  \\  
    p(-) &= e^{- \beta h_b } \left[ e^{\beta C} + e^{\beta D} \right]^{\Nb} / \Zmf \textrm{ ,}       
\end{align}   
related by $p(+) + p(-) = 1$.   
As usual, the important physical observables of the system can be calculated from the derivatives of the logarithm of the partition function (\ref{eq:Zmf}), such as the bulk magnetization $m$ (per spin),  
\begin{equation} \label{eq:m0Definition} 
    m = \frac{1}{\Nb} \sum_{ i }^{ \Nb } \frac{ \sum_{ \bm{\sigma} } \left[ \sigma_i e^{ - \beta \Hmf } \right] }{ \Zmf } 
    = \frac{ \sum_{ \bm{\sigma} } \left[ \sum_{i}^{\Nb} \sigma_i e^{ - \beta \Hmf } \right] }{ \Nb \Zmf } \textrm{ .}             
\end{equation} 
Noticing that $(1/\beta) \cdot (\partial \Zmf /\partial h) = \sum_{ \bm{\sigma} } \left[ \sum_{i}^{\Nb} \sigma_i e^{ - \beta \Hmf } \right]$, it is straightforward to show that 
\begin{equation} \label{eq:m0Intermediate} 
    m = \frac{1}{\beta \Nb} \frac{\partial \ln[\Zmf]}{\partial h} \textrm{ .}    
\end{equation} 
Performing the derivative in Eq. (\ref{eq:m0Intermediate}) we find the mean field solution for the average bulk magnetization,  
\begin{equation} \label{eq:m0mf}   
    m = p(+) \frac{e^{\beta A} - e^{\beta B}}{e^{\beta A} + e^{\beta B}} 
        + p(-) \frac{ e^{\beta C} - e^{\beta D} }{ e^{\beta C} + e^{\beta D} } \textrm{ .}     
\end{equation} 
This is a self-consistent, transcendental equation for $m$, since the exponential terms $A$, $B$, $C$ and $D$ depend on $m$. It does not seem possible to find a closed form expression for $m$. 
Clearly, the first term corresponds to the average bulk magnetization when the super spin is up ($p(+)=1$), while the second one is the average bulk magnetization when the super spin is down ($p(-)=1$).       
Eq. (\ref{eq:m0mf}) can be approximated, however, by simplifying the expression for the partition function (\ref{eq:Zmf}) in the thermodynamic limit. 
First, notice that, 
\begin{align*} 
    \Zmf & \propto  
    \left[ e^{\beta A} + e^{\beta B} \right]^{\Nb}    
    + \left[ e^{\beta C} + e^{\beta D} \right]^{\Nb} = \\   
\refstepcounter{equation}\tag{\theequation}    &=  \left[ e^{\beta A} + e^{\beta B} \right]^{\Nb} \cdot \left( 1 + \left[ (e^{\beta C} + e^{\beta D})/(e^{\beta A} + e^{\beta B}) \right]^{\Nb} \right)         
    \textrm{ ,}  
\end{align*} 
therefore, if 
\begin{equation} \label{eq:conditionSuperUp}      
    e^{\beta A} + e^{\beta B} > e^{\beta C} + e^{\beta D} \textrm{ ,}   
\end{equation}
the ratio $ (e^{\beta C} + e^{\beta D})/(e^{\beta A} + e^{\beta B}) < 1$, and in the limit $\Nb \rightarrow \infty$, the term $ \left[ (e^{\beta C} + e^{\beta D})/(e^{\beta A} + e^{\beta B}) \right]^{\Nb}$ goes to zero. Consequently,  
\begin{equation} \label{eq:ZmfSuperUp}  
    \Zmf \approx e^{- \beta h_b } \left[ e^{\beta A} + e^{\beta B} \right]^{\Nb} \textrm{ .}    
\end{equation}   
Applying this approximation into Eq. (\ref{eq:m0Intermediate}) yields 
\begin{equation} \label{eq:m0SuperUp} 
    m \approx \frac{e^{\beta A} - e^{\beta B}}{e^{\beta A} + e^{\beta B}} 
    = \tanh \left[ \beta h_{+} \right] \textrm{ ,}   
\end{equation}      
where in the last step we cancelled out the common term $e^{\beta \hsrd}$. Interestingly, comparing result (\ref{eq:m0SuperUp}) with Eq. (\ref{eq:m0mf}) implies $p(+) = 1$ and $p(-) = 0$, i.e. in the thermodynamic limit, the super spin is expected to be always aligned with its local field $h_s$ if condition (\ref{eq:conditionSuperUp}) is satisfied.           
On the other hand, if instead  
\begin{equation} \label{eq:conditionSuperDown}  
     e^{\beta A} + e^{\beta B} < e^{\beta C} + e^{\beta D} \textrm{ ,}       
\end{equation} 
we have  
\begin{equation} \label{eq:ZmfSuperDown}    
    \Zmf \approx e^{- \beta h_b } \left[ e^{\beta C} + e^{\beta D} \right]^{\Nb} \textrm{ .}   
\end{equation}     
and           
\begin{equation} \label{eq:m0SuperDown}  
    m \approx \frac{ e^{\beta C} - e^{\beta D} }{ e^{\beta C} + e^{\beta D} } 
    = \tanh \left[ \beta h_{-} \right] \textrm{ .}       
\end{equation} 
This result, in turn, implies that $p(-) = 1$ and $p(+) = 0$, namely the super spin is expected to be always anti-aligned with its local field in the thermodynamic limit if condition (\ref{eq:conditionSuperDown}) is satisfied instead.      
   
In conclusion, to find the solutions of $m$ for fixed values of the external parameters, one has to solve Eqs. (\ref{eq:m0SuperUp}) and (\ref{eq:m0SuperDown}) depending if temperature $T$ and average bulk magnetization $m$ satisfy, respectively, conditions (\ref{eq:conditionSuperUp}) or (\ref{eq:conditionSuperDown}). The behavior of the bulk, according to the mean field description in the thermodynamic limit, is consistent with a ferromagnetic system with $\Nb$ fully-connected spins, each pair interacting with strength $J = \Jrd / (\Nb - 1)$, under the influence of an effective "external" field $( \pm J_s + h )$. The sign of $J_s$ depends on the state of the super spin which, as a function of $T$ and $m$, may flip, acting like a switch. This is a consequence of the cooperative behavior emerging from the interaction between the super spin and the bulk.         
           
From the partition function (\ref{eq:Zmf}) one can also calculate the mean field expression for the free energy, 
\begin{equation} \label{eq:FmfDefinition}    
    \Fmf = - k_b T \ln[ \Zmf ] \textrm{ .}    
\end{equation} 
Using the respective $\Nb \rightarrow \infty$ approximations (\ref{eq:ZmfSuperUp}) and (\ref{eq:ZmfSuperDown}), the (intensive) free energy is given by  
\begin{equation} \label{eq:Fmf}  
    \frac{ \Fmf }{ \Nb } \propto \Jrd \frac{ m^2 }{ 2 } - \hsrd - \ln \left[ \cosh \left( \beta \cdot ( \pm J_s + h + \Jrd m ) \right) \right] / \beta    \textrm{ ,}   
\end{equation} 
where $+$ or $-$ hold if condition (\ref{eq:conditionSuperUp}) or (\ref{eq:conditionSuperDown}) is satisfied, respectively.    
 
For the bulk parameters, we define $h = -1$ and $J = 1 / ( \Nb - 1 ) $. Notice that these definitions, i.e. $h$ independent of $\Nb$ and $J \propto \Nb ^ {-1} $, are important to have the Hamiltonian $H \propto \Nb$ and the free energy $F \propto \Nb$, and therefore well-defined extensive energies \cite{castellani_spin-glass_2005}. For the super spin parameters, we define $h_s = \Nb$ to achieve field neutrality, i.e. $ \hsrd = h_s / \Nb = |h| = 1 $. For the interaction constant, first notice that each bulk spin interacts with $\Nb - 1$ other bulk spins with strength $J$, therefore, if all other spins are aligned, the energy contribution of the bulk interactions is of the order $J \cdot ( \Nb - 1 ) = \Jrd = 1$. We set $J_s = 1.01$, in order to study a regime where the super spin interacts with the bulk with a slightly stronger interaction than the bulk interactions, even when global opinion accordance is achieved within the bulk.   

\begin{figure*}[!htb]   
    \centering
    \includegraphics[width=\linewidth]{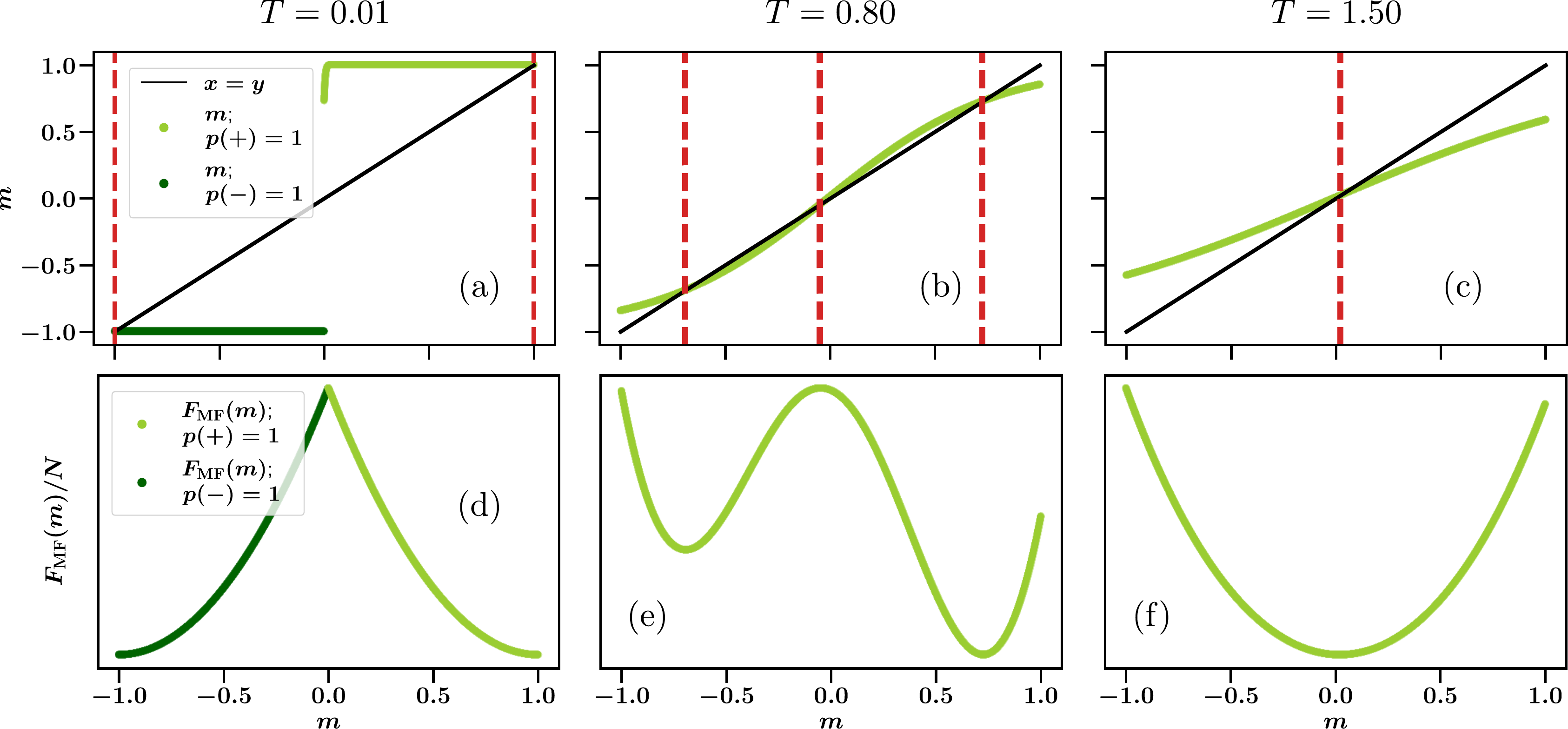}
    \caption{\textbf{Mean field solution.} Graphical solutions of the mean field self-consistent equation for for the average bulk magnetization $m$ (a-c), Eqs. (\ref{eq:m0SuperUp}) and (\ref{eq:m0SuperDown}), and the free energy $\Fmf$ (\ref{eq:Fmf}) as a function of $m$ (d-f), for three values of $ T = \{ 0.01 , 0.80, 1.50 \} $, in the thermodynamic limit $\Nb \rightarrow \infty$. The light and dark green curves correspond to the regimes where conditions (\ref{eq:conditionSuperUp}) and (\ref{eq:conditionSuperDown}) are satisfied, respectively, and therefore the regimes where the super spin is predicted to be always up ($p(+) = 1$) or down ($p(-1) = 1$) in the thermodynamic limit. The vertical red dashed lines in (a-c) indicate the mean-field solutions of $m$ (both stable and unstable ones), for which Eqs. (\ref{eq:m0SuperUp}) or (\ref{eq:m0SuperDown}) intersect the bisector $ x = y $. The parameters are: $h = -1$, $\Jrd = 1 $, $\hsrd = 1 $, $J_s = 1.01 $.} 
    \label{fig:mean-field-solution}    
\end{figure*}

In Fig. (\ref{fig:mean-field-solution}) we plot the mean-field solutions for $m$, Eqs. (\ref{eq:m0SuperUp}) and (\ref{eq:m0SuperDown}), as well as the free energy (\ref{eq:Fmf}), for three values of the temperature $ T = \{ 0.01 , 0.80, 1.50 \} $, in the thermodynamic limit $ \Nb \rightarrow \infty $.         
For low temperature, there are two solutions for $m$ (Fig. (\ref{fig:mean-field-solution})a) with similar free energy (Fig. (\ref{fig:mean-field-solution})d), corresponding to consensus aligned with the bulk field $h$, $m \approx -1$, and opposite to it, $m \approx 1$. In this regime, the super spin is expected to always be aligned with the bulk, as seen from the dark green curve (super spin always down, anti-aligned with its field), which intersects the bisector $x = y$ at $m \approx -1$, while the light green one (super spin always up, aligned with its field) intersects the bisector at $m \approx 1$.  
At intermediate temperatures, the situation is quite different, as, independently of the bulk average magnetization, the super spin is expected to always be up, aligned with its field $h_s$. The solutions of $m$ at positive and negative magnetization are now $ | m | < 1 $ (Fig. (\ref{fig:mean-field-solution})b), indicating that consensus is no longer stable, as expected from a more liberal social climate. A new solution also appears near $m \approx 0$, but the respective free energy plot (Fig. (\ref{fig:mean-field-solution})e) indicates that this solution is unstable.  
Finally, at high temperatures, the system has only a single stable solution at $m \approx 0$ (Fig. (\ref{fig:mean-field-solution})c and f), which indicates indicates a social climate where opinions are highly varied and any kind of consensus is unstable. The super spin, as before, is expected to be always aligned with its field $h_s$. 
    
\section{Numerical results}   
\label{sec:Numerical-results}   

Next, we will study finite systems through the numerical estimation of the average bulk $m$ and super spin $m_s$ magnetization from Monte Carlo simulations as a function of the temperature $T$, and compare with the mean field predictions (\ref{eq:m0SuperUp}) and (\ref{eq:m0SuperDown}), depending on which condition (\ref{eq:conditionSuperUp}) or (\ref{eq:conditionSuperDown}) is satisfied.  
        
For the Monte Carlo simulations, we use the Metropolis algorithm \cite{bottcher_computational_2021}. One Monte Carlo step corresponds to a full lattice sweep. For each temperature $T$ and a given initial spin configuration, we thermalize by disregarding the first $ \Ntherm = 2 \cdot 10^4 $ steps, and then average over $ \NMC = 10^4 $ spin configurations, storing spin configurations for averaging only every two steps, to reduce correlations between configurations. We repeat this procedure and average over $ \NIC = 10^4 $ independent initial spin configurations.        
For low $T$, energy barriers between metastable states might be large compared to the thermal energy \cite{castellani_spin-glass_2005}. In finite systems, it is understood that there exists always a finite time for the system to overcome these barriers and reach thermal equilibrium, but this time might be much longer than the Monte Carlo simulation times considered.  
A possible approach to check if we are averaging the bulk magnetization $m$ over states which have different time averaged magnetization, when considering several independent initial spin configuration, is to split the measure of the average $m$ into two components \cite{castellani_spin-glass_2005},  
\begin{equation} \label{eq:m0Split}         
    m = g_{+} m_{+} + g_{-} m_{-} \textrm{ ,}  
\end{equation} 
where $m_{+}$ is the bulk magnetization averaged over several independent initial spin configurations which end up with a positive time-averaged bulk magnetization, while $m_{-}$ is the average over initial spin configurations which end up in turn with a negative time-averaged bulk magnetization. $ g_{+} $ and $ g_{-} $ correspond to the fraction of independent initial configurations that
end up with either positive or negative average bulk magnetization, respectively. Naturally, they are constrained by the normalization condition $ g_{+} + g_{-} = 1 $. We split the average super spin magnetization in an analogous way, 
\begin{equation} \label{eq:msSplit} 
    m_s = g_{s+} m_{s+} + g_{s-} m_{s-} \textrm{ .}    
\end{equation}    
where the split components $m_{s+}$ and $m_{s-}$ refer now to the average super spin magnetization over several independent initial configurations when its time average ends up positive or negative, respectively, and $g_{s+}$ ($g_{s-}$) is the fraction of initial configurations that end up with positive (negative) average super spin magnetization. 
                       
\begin{figure*}           
    \centering
    \includegraphics[width=0.80\linewidth]{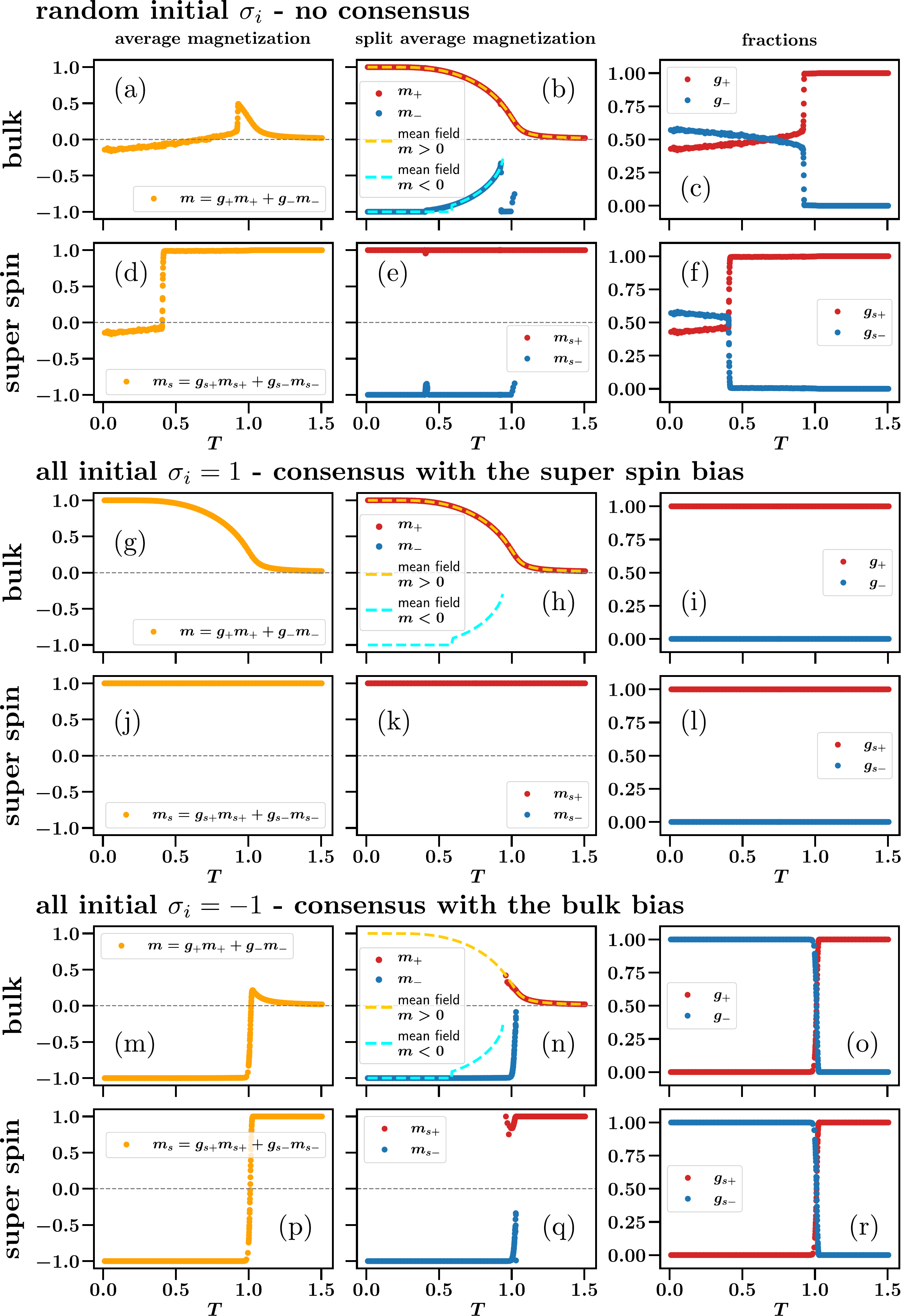}     
    \caption{\textbf{Numerical results.} Results from Monte Carlo simulations of the average bulk $m$ and super spin $m_s$ magnetization (first column), their partition into positive or negative average magnetization (middle column) and the fractions of Monte Carlo initial configurations which evolve towards a state with positive or negative average magnetization (last column), as a function of $T \in \left[0.01,1.50\right]$, for $N \sim 10^4$ and three different initial conditions: no consensus (uniform distribution of $\{ \sigma_i \}$ and random $\sigma_s$, (a-f)), consensus with the super spin bias (all $\{ \sigma_i \}, \sigma_s = 1$, (g-l)), and consensus with the bulk bias (all $ \{ \sigma_i \}, \sigma_s = -1$, (m-r)). For each point in $T$, results are averages over $ \NIC = 10^4 $ independent Monte Carlo simulations. In plots (b), (h) and (n) we also show the mean-field solutions for $m$ in the thermodynamic limit (colored dashed lines), as predicted by Eqs. (\ref{eq:m0SuperUp}) and (\ref{eq:m0SuperDown}) \mycomment{, corresponding to the minima of the free energy (\ref{eq:Fmf})}. The horizontal grey dashed lines indicate $m=0$ or $m_s=0$. The parameters are: $h = -1$, $\Jrd = 1 $, $\hsrd = 1 $, $J_s = 1.01 $.}
    \label{fig:numerical-results-N10000} 
\end{figure*}

In Fig. (\ref{fig:numerical-results-N10000}) we plot the values of both $m$ and $m_s$, as well as their respective split components, as defined in Eqs. (\ref{eq:m0Split}) and (\ref{eq:msSplit}), as a function of temperature $T$, for a system with $N \sim 10^4$ spins, for three different initial conditions: starting with a uniform distribution of $\sigma_i$ (no consensus), with all spins $\sigma_i = 1$, aligned with the super spin field $h_s$ (consensus with the bias of the leader), and all spins $\sigma_i = -1$, aligned with the bulk field $h$ (consensus with the bulk bias), with the same respective initial conditions applied to the super spin $\sigma_s$.         
       
Looking first at the case starting with all $\sigma_i$ randomly assigned (Fig. (\ref{fig:numerical-results-N10000})a-f), a configuration corresponding to no consensus, the average bulk magnetization (Fig. (\ref{fig:numerical-results-N10000})a) exhibits different behavior depending on the temperature $T$. For low $T \lesssim 1$, the system can evolve towards one of two stable states (Fig. (\ref{fig:numerical-results-N10000})b), with either positive $m_{+}$ or negative $m_{-}$ average bulk magnetization. The numerical values of $m_{+}$ and $m_{-}$ are consistent with the mean field solutions for $m$ (dashed lines in Fig. (\ref{fig:numerical-results-N10000})b), where $m_{+}$ is predicted to be the global minimum of the free energy (see Fig.(\ref{fig:mean-field-solution})b) while $m_{-}$ is a local one. Despite this, numerical results show that the state $m_{-}$ is a persistent metastable state for low $T$ and, curiously, has a higher probability for the system to evolve to, as seen from Fig. (\ref{fig:numerical-results-N10000})c, where $g_{-} > g_{+}$ for low $T$. The mixed average of the metastable $m<0$ and stable $m>0$ states with corresponding uneven weights $g_{-}$ and $g_{+}$ therefore leads to a negative value in the average bulk magnetization seen in Fig. (\ref{fig:numerical-results-N10000})a.  
This interestingly suggests that in an authoritarian regime, starting from a state with no consensus, it is more likely that the bulk reaches consensus with the opinion of the peers and the leader adapts to the majority choice (Fig. (\ref{fig:numerical-results-N10000})f).   
For $ T \gtrsim 1 $, the system undergoes a phase transition, and only the positive $m_{+}$ state is observed, as $g_{+} = 1$ in Fig. (\ref{fig:numerical-results-N10000})c, again consistent with the mean field solution (see Fig.(\ref{fig:mean-field-solution})c and f), leading to a positive $m$ in Fig. (\ref{fig:numerical-results-N10000})a. 
Considering the average super spin magnetization $m_s$ (Fig. (\ref{fig:numerical-results-N10000})d), for low $T \lesssim 0.4$, two states can be observed, with either positive $m_{s+}$ or negative $m_{s-}$ super spin magnetization (Fig. (\ref{fig:numerical-results-N10000})e), with the negative one $m_{s-}$ being again more likely to be observed for low $T$, as $g_{s-} > g_{s+}$ in Fig. (\ref{fig:numerical-results-N10000})f. For higher $ T \gtrsim 0.4 $, only the positive $m_{s+}$ is observed, with $g_{s+} \approx 1$ in Fig. (\ref{fig:numerical-results-N10000})f.
Notice that this occurs for a much lower temperature than the phase transition observed in the bulk magnetization, suggesting that more fluctuations are necessary to observe the change in opinion for the bulk. Notice also that, except for $ T \approx 0.4$, $m_{s+} = 1$ and $m_{s-} = - 1$, meaning that the super spin is either always up or down depending on which state the system evolves to, consistently with the mean field prediction in the thermodynamic limit.            
       
For the case of initial configurations with all $\sigma_i = 1$ (Fig. (\ref{fig:numerical-results-N10000})g-l), a configuration corresponding with consensus with the super spin bias, the long-time behavior of the system is quite different. Now, only the state with positive average bulk magnetization $m_s$ is observed (Fig. (\ref{fig:numerical-results-N10000})h and i), corresponding to the global free energy minimum (dashed line in Fig. (\ref{fig:numerical-results-N10000})h). Consequently, the average bulk magnetization (Fig. (\ref{fig:numerical-results-N10000})g) no longer changes sign with temperature.
For this case, the super spin is always up (see Fig. (\ref{fig:numerical-results-N10000})j-l), aligned with its field $h_s$.        
     
Lastly, for the case with spins starting all in the $\sigma_i = -1$ state, corresponding to consensus with the bulk bias, the behavior is again quite different. The average bulk magnetization (Fig. (\ref{fig:numerical-results-N10000})m-o) is negative for low $T \lesssim 1$, with $m = -1$ indicating that all bulk spins are aligned with the bulk field $h$. For $T \gtrsim 1$, the state with negative $m$ becomes unstable, and the average $m$ becomes positive even when starting with all $\sigma_i = -1$. The average super spin magnetization $m_s$ (Fig. (\ref{fig:numerical-results-N10000})p-r) also changes sign with temperature. For low $T \lesssim 1$, it remains aligned with the bulk field $h$ ($m_s = -1$), while for high $T \gtrsim 1$ it flips, remaining aligned with its field $h_s$ ($m_s = 1$). Near the phase transition $T \approx 1$, the state of the super spin fluctuates over time, as evidenced by the value of $| m_s | < 1$ (Fig. (\ref{fig:numerical-results-N10000})p and q). Interestingly, unlike the case starting from a configuration with no consensus (Fig. (\ref{fig:numerical-results-N10000})a-f), the transition temperature for the super spin is similar to that of the bulk, at $ T \approx 1$. It is also noteworthy that the numerical value for the partial average magnetization $m_{-}$ (Fig. (\ref{fig:numerical-results-N10000})n) remains close to $m_{-} \approx -1$ until $T \approx 1$, despite not being a minimum of the mean field free energy (blue dashed line).  
Similar results are obtained if we consider only nearest neighbour interactions in the bulk (see supplementary Fig. S1), where the effect of the changing sign of $m$ with $T$ seems to become less relevant as the system size increases (see supplementary Fig. S2).  
    
\section{Conclusions}   
\label{sec:Conclusions}

We have studied the opinion dynamics in a social system under the influence of a charismatic leader by means of a generalized Ising model where all spins interact with a super spin via strong couplings. The interesting feature of the model is the appearance of a change in sign of both the bulk and leader average opinions as function of the social climate, behavior not usually found in Ising models. 
In summary, the mean-field approach (Fig. (\ref{fig:mean-field-solution})) suggests that the state of total ($m=1$) or partial ($0<m<1$) consensus with the leader is predicted to be the stable state, for any social climate $T$. 
For finite systems, numerical results (Fig. (\ref{fig:numerical-results-N10000})) depend on the initial conditions for a totalitarian regime (low $T$) due to metastable states corresponding to total ($m=-1$) or partial ($-1<m<0$) consensus with the bulk bias. When there is disagreement with the leader, i.e. no consensus ($m\approx0$) or consensus with the bulk ($m=-1$), the sign of the time-averaged social mood $m$ depends on the social climate $T$. If the system starts from a state of no consensus ($m \approx 0$, Fig. (\ref{fig:numerical-results-N10000})a-f), the effect that the social mood $m$ changes sign as function of temperature is due to an ensemble average over both metastable and stable states. Interestingly, the probabilities to evolve to either state are unequal (Fig. (\ref{fig:numerical-results-N10000})c), with the metastable state being more probable than the stable one for low $T$. For a more liberal regime $T \gtrsim 1$, only the stable state of partial consensus with the leader ($m>0$) is observed (Fig. (\ref{fig:numerical-results-N10000})b-c) in the long-time behavior of the system. These results suggest that, if a social group is initially in a state of no consensus, agreement with the ideology of a leader is difficult, because it is more likely for the bulk to align its opinion with its own bias, when the social climate is authoritarian (low $T$). For the initial condition corresponding to consensus with the bulk bias ($m=-1$, Fig. (\ref{fig:numerical-results-N10000})m-q), $m$ changes sign with $T$ because the system always remains stuck in a persistent metastable state, aligned with the bulk ideology ($m=-1$), and, as the social climate changes to a more liberal regime, the system transitions into the equilibrium state of partial agreement with the leader ($m>0$). This behaviour is also observed if only nearest-neighbor interactions are considered in the bulk, corresponding to a scenario of limited social contact within the social group. 
Possible future directions for the study of this model include analyzing how the system behaves as a function of the super spin parameters $h_s$ and $J_s$ and investigating non-equilibrium properties, such as the transition rates, as is usually done in the context of opinion dynamics \cite{castellano_statistical_2009,vilela_majority-vote_2019,redner_reality-inspired_2019,baldassarri_ising_2023,corberi_kinetics_2023,helfmann_modelling_2023,meyer_time_2024}.

\section*{CRediT authorship contribution statement}  

\textbf{Tiago S. A. N. Simões}: Software, Investigation, Formal analysis, Writing - Original Draft, Writing - Review \& Editing.  
\textbf{Antonio Coniglio}: Formal analysis, Validation, Writing - Original Draft, Writing - Review \& Editing.  
\textbf{Hans J. Herrmann}: Conceptualization, Methodology, Validation, Writing - Original Draft, Writing - Review \& Editing.  
\textbf{Lucilla de Arcangelis}: Conceptualization, Methodology, Validation, Writing - Original Draft, Writing - Review \& Editing.   
  
\section*{Declaration of competing interest}  
The authors declare that they have no known competing financial interests or personal relationships that could have appeared to influence the work reported in this paper. 
  
\section*{Acknowledgments}  
 
L.d.A. acknowledges support from the Italian MUR project PRIN2017WZFTZP and from PRIN 2022 PNRR P202247YKL. H.J.H. thanks FUNCAP and the INCT-SC for financial support.

\appendix

\section{Derivation of the mean field partition function} 
\label{section:ZmfDerivation} 
               
To derive the mean field partition function, we start by applying the expression for the mean field Hamiltonian $ \Hmf $ (\ref{eq:Hmf}) in the formula $ \Zmf = \sum_{ \bm{\sigma} } e^{- \beta \Hmf } $, 
\begin{align*}   
\refstepcounter{equation}\tag{\theequation}     \Zmf = \prodB \left( \sumStates \right) \sum_{ \sigma_s = \pm 1 }^{} \left[ e^{ - \beta \Hmf } \right]    
         \propto \prodB \left( \sumStates \right) \sum_{ \sigma_s = \pm 1 }^{} 
         \left[ e^{ \beta \cdot \left( h_s \sigma_s + J_s \sigma_s + h + J m \cdot (\Nb - 1 ) \right) \sumB \sigma_i}  \right] 
      \textrm{ ,}   
\end{align*}   
where we will ignore for now the term $e^{- \beta h_b }$.   
Expanding the sum $ \sum_{ \sigma_s = \pm 1 }^{} $ over the super spin states we get  
\begin{align*}   
\refstepcounter{equation}\tag{\theequation}    \Zmf \propto  e^{ \beta h_s } \prodB \left( \sumStates \right) e^{ \beta \cdot \left( h + J_s + J m \cdot (\Nb - 1 ) \right) \sumB \sigma_i} +  e^{ -\beta h_s } \prodB \left( \sumStates \right) e^{ \beta \cdot \left( h - J_s + J m \cdot (\Nb - 1 ) \right) \sumB \sigma_i}  \textrm{ .}  \label{eq:intermediateZmf}                   
\end{align*} 
For brevity of notation, we will define $ h_{+} \equiv h + J_s + J m \cdot ( \Nb - 1 ) $ and $ h_{-} \equiv h - J_s + J m \cdot ( \Nb - 1 ) $, which can be interpreted as the mean field effective fields felt by the bulk spins due to the field $h$, the interactions with all other $\Nb - 1$ bulk spins, and the interaction with the super spin, depending if the latter is in the up-state ($h_{+}$) or down-state ($h_{-}$). Applying these definitions in (\ref{eq:intermediateZmf}) simplifies the expression to         
\begin{align*} \label{eq:ZmfWithPMfields}   
\refstepcounter{equation}\tag{\theequation}    \Zmf \propto e^{ \beta h_s } \prodB \left( \sumStates \right) e^{ \beta h_{+} \sumB \sigma_i } + e^{ -\beta h_s } \prodB \left( \sumStates \right) e^{ \beta h_{-} \sumB \sigma_i }  \textrm{ .}        
\end{align*}   
To proceed, notice that  
\begin{align*} 
    & \prodB \left( \sumStates \right) e^{ \beta h_{+} \sumB \sigma_i } = \prodB \left( \sumStates \right) e^{ \beta h_{+} \sigma_1 } e^{ \beta h_{+} \sigma_2 } \cdots e^{ \beta h_{+} \sigma_N } = \\ 
\refstepcounter{equation}\tag{\theequation}    & = \sum_{ \sigma_1 = \pm 1 }^{} e^{ \beta h_{+} \sigma_1 } \sum_{ \sigma_2 = \pm 1 }^{} e^{ \beta h_{+} \sigma_2 } \cdots \sum_{ \sigma_N = \pm 1 }^{} e^{ \beta h_{+} \sigma_N } = \prodB \left( e^{ \beta h_{+} } + e^{ - \beta h_{+} } \right) = \left( e^{ \beta h_{+} } + e^{ - \beta h_{+} } \right)^{ \Nb } \textrm{ ,}           
\end{align*} 
where the second to last step is possible since all bulk spins are identical. Using the same logic for the exponential term with $ h_{-} $, Eq. (\ref{eq:ZmfWithPMfields}) is equal to   
\begin{align*}  
    \Zmf & \propto e^{ \beta h_s } \cdot \left[ e^{ \beta h_{+} } + e^{ - \beta h_{+} } \right]^{ \Nb } + e^{ -\beta h_s } \cdot \left[ e^{ \beta h_{-} } + e^{ - \beta h_{-} } \right]^{ \Nb } = \\ 
\refstepcounter{equation}\tag{\theequation}    & = \left[ e^{ \beta \cdot (\frac{h_s}{\Nb} + h_{+} ) } + e^{ \beta \cdot ( \frac{h_s}{\Nb} - h_{+} ) } \right]^{ \Nb } + \left[ e^{ \beta \cdot ( - \frac{h_s}{\Nb} + h_{-} ) } + e^{ \beta \cdot ( - \frac{h_s}{\Nb} - h_{-} ) } \right]^{ \Nb } \textrm{ .}    
\end{align*}  
Introducing the following definitions,   
\begin{align} 
    A &\equiv \frac{h_s}{\Nb} + h_{+} = \hsrd + J_s + h + \Jrd m \textrm{ ,}  \\   
    B &\equiv \frac{h_s}{\Nb} - h_{+} = \hsrd - J_s - h - \Jrd m \textrm{ ,}  \\    
    C &\equiv - \frac{h_s}{\Nb} + h_{-} = - \hsrd - J_s + h + \Jrd m \textrm{ ,} \\    
    D &\equiv - \frac{h_s}{\Nb} - h_{-} = - \hsrd + J_s - h - \Jrd m \textrm{ ,}   
\end{align}   
where $ \hsrd \equiv h_s / \Nb $ and $ \Jrd \equiv ( \Nb - 1 ) J $, and retrieving the dropped out term $e^{- \beta h_b }$, we obtain the expression for the mean field partition function as reported in the main text,  
\begin{equation}
    \Zmf = e^{- \beta h_b } \cdot \left( 
    \left[ e^{\beta A} + e^{\beta B} \right]^{\Nb}   
    + \left[ e^{\beta C} + e^{\beta D} \right]^{\Nb} \right) \textrm{ .}      
\end{equation}

\bibliographystyle{elsarticle-num} 
\bibliography{references} 

\begin{thebibliography}{10}
\expandafter\ifx\csname url\endcsname\relax
  \def\url#1{\texttt{#1}}\fi
\expandafter\ifx\csname urlprefix\endcsname\relax\def\urlprefix{URL }\fi
\expandafter\ifx\csname href\endcsname\relax
  \def\href#1#2{#2} \def\path#1{#1}\fi

\bibitem{galam_sociophysics_1982}
S.~Galam, Y.~Gefen~(Feigenblat), Y.~Shapir, Sociophysics: {{A}} new approach of sociological collective behaviour. {{I}}. mean-behaviour description of a strike, The Journal of Mathematical Sociology 9~(1) (1982) 1--13.
\newblock \href {https://doi.org/10.1080/0022250X.1982.9989929} {\path{doi:10.1080/0022250X.1982.9989929}}.

\bibitem{jones_simulation_1985}
F.~Jones, Simulation {{Models}} of {{Group Segregation}}, The Australian and New Zealand Journal of Sociology 21~(3) (1985) 431--444.
\newblock \href {https://doi.org/10.1177/144078338502100307} {\path{doi:10.1177/144078338502100307}}.

\bibitem{castellano_statistical_2009}
C.~Castellano, S.~Fortunato, V.~Loreto, Statistical physics of social dynamics, Reviews of Modern Physics 81~(2) (2009) 591--646.
\newblock \href {https://doi.org/10.1103/RevModPhys.81.591} {\path{doi:10.1103/RevModPhys.81.591}}.

\bibitem{weidlich_statistical_1971}
W.~Weidlich, The {{Statistical Description}} of {{Polarization Phenomena}} in {{Society}}, British Journal of Mathematical and Statistical Psychology 24~(2) (1971) 251--266.
\newblock \href {https://doi.org/10.1111/j.2044-8317.1971.tb00470.x} {\path{doi:10.1111/j.2044-8317.1971.tb00470.x}}.

\bibitem{baldassarri_ising_2023}
S.~Baldassarri, A.~Gallo, V.~Jacquier, A.~Zocca, Ising model on clustered networks: {{A}} model for opinion dynamics, Physica A: Statistical Mechanics and its Applications 623 (2023) 128811.
\newblock \href {https://doi.org/10.1016/j.physa.2023.128811} {\path{doi:10.1016/j.physa.2023.128811}}.

\bibitem{schelling_dynamic_1971}
T.~C. Schelling, Dynamic models of segregation, The Journal of Mathematical Sociology 1~(2) (1971) 143--186.
\newblock \href {https://doi.org/10.1080/0022250X.1971.9989794} {\path{doi:10.1080/0022250X.1971.9989794}}.

\bibitem{axelrod_dissemination_1997}
R.~Axelrod, The {{Dissemination}} of {{Culture}}: {{A Model}} with {{Local Convergence}} and {{Global Polarization}}, Journal of Conflict Resolution 41~(2) (1997) 203--226.
\newblock \href {https://doi.org/10.1177/0022002797041002001} {\path{doi:10.1177/0022002797041002001}}.

\bibitem{clifford_model_1973}
P.~Clifford, A.~Sudbury, A model for spatial conflict, Biometrika 60~(3) (1973) 581--588.
\newblock \href {https://doi.org/10.1093/biomet/60.3.581} {\path{doi:10.1093/biomet/60.3.581}}.

\bibitem{holley_ergodic_1975}
R.~A. Holley, T.~M. Liggett, Ergodic {{Theorems}} for {{Weakly Interacting Infinite Systems}} and the {{Voter Model}}, The Annals of Probability 3~(4) (1975) 643--663.
\newblock \href {https://doi.org/10.1214/aop/1176996306} {\path{doi:10.1214/aop/1176996306}}.

\bibitem{redner_reality-inspired_2019}
S.~Redner, Reality-inspired voter models: {{A}} mini-review, Comptes Rendus Physique 20~(4) (2019) 275--292.
\newblock \href {https://doi.org/10.1016/j.crhy.2019.05.004} {\path{doi:10.1016/j.crhy.2019.05.004}}.

\bibitem{sznajd-weron_opinion_2000}
K.~{Sznajd-Weron}, J.~Sznajd, Opinion evolution in closed community, International Journal of Modern Physics C 11~(06) (2000) 1157--1165.
\newblock \href {https://doi.org/10.1142/S0129183100000936} {\path{doi:10.1142/S0129183100000936}}.

\bibitem{sznajd-weron_sznajd_2005}
K.~{Sznajd-Weron}, Sznajd model and its applications (Mar. 2005).
\newblock \href {http://arxiv.org/abs/physics/0503239} {\path{arXiv:physics/0503239}}, \href {https://doi.org/10.48550/arXiv.physics/0503239} {\path{doi:10.48550/arXiv.physics/0503239}}.

\bibitem{de_oliveira_isotropic_1992}
M.~J. {de Oliveira}, Isotropic majority-vote model on a square lattice, Journal of Statistical Physics 66~(1) (1992) 273--281.
\newblock \href {https://doi.org/10.1007/BF01060069} {\path{doi:10.1007/BF01060069}}.

\bibitem{vilela_majority-vote_2019}
A.~L.~M. Vilela, C.~Wang, K.~P. Nelson, H.~E. Stanley, Majority-vote model for financial markets, Physica A: Statistical Mechanics and its Applications 515 (2019) 762--770.
\newblock \href {https://doi.org/10.1016/j.physa.2018.10.007} {\path{doi:10.1016/j.physa.2018.10.007}}.

\bibitem{mobilia_does_2003}
M.~Mobilia, Does a {{Single Zealot Affect}} an {{Infinite Group}} of {{Voters}}?, Physical Review Letters 91~(2) (2003) 028701.
\newblock \href {https://doi.org/10.1103/PhysRevLett.91.028701} {\path{doi:10.1103/PhysRevLett.91.028701}}.

\bibitem{yildiz_discrete_2011}
E.~Yildiz, D.~Acemoglu, A.~E. Ozdaglar, A.~Saberi, A.~Scaglione, Discrete {{Opinion Dynamics}} with {{Stubborn Agents}} (Jan. 2011).
\newblock \href {https://doi.org/10.2139/ssrn.1744113} {\path{doi:10.2139/ssrn.1744113}}.

\bibitem{verma_impact_2014}
G.~Verma, A.~Swami, K.~Chan, The impact of competing zealots on opinion dynamics, Physica A: Statistical Mechanics and its Applications 395 (2014) 310--331.
\newblock \href {https://doi.org/10.1016/j.physa.2013.09.045} {\path{doi:10.1016/j.physa.2013.09.045}}.

\bibitem{khalil_zealots_2018}
N.~Khalil, M.~San~Miguel, R.~Toral, Zealots in the mean-field noisy voter model, Physical Review E 97~(1) (2018) 012310.
\newblock \href {https://doi.org/10.1103/PhysRevE.97.012310} {\path{doi:10.1103/PhysRevE.97.012310}}.

\bibitem{mobilia_role_2007}
M.~Mobilia, A.~Petersen, S.~Redner, On the role of zealotry in the voter model, Journal of Statistical Mechanics: Theory and Experiment 2007~(08) (2007) P08029.
\newblock \href {https://doi.org/10.1088/1742-5468/2007/08/P08029} {\path{doi:10.1088/1742-5468/2007/08/P08029}}.

\bibitem{klamser_zealotry_2017}
P.~P. Klamser, M.~Wiedermann, J.~F. Donges, R.~V. Donner, Zealotry effects on opinion dynamics in the adaptive voter model, Physical Review E 96~(5) (2017) 052315.
\newblock \href {https://doi.org/10.1103/PhysRevE.96.052315} {\path{doi:10.1103/PhysRevE.96.052315}}.

\bibitem{toral_finite_2007}
R.~Toral, C.~J. Tessone, Finite size effects in the dynamics of opinion formation, Communications in Computational Physics 2 (2007) 177--195.

\bibitem{ritchie_population_2023}
H.~Ritchie, L.~{Rod{\'e}s-Guirao}, E.~Mathieu, M.~Gerber, E.~{Ortiz-Ospina}, J.~Hasell, M.~Roser, Population {{Growth}}, https://ourworldindata.org/population-growth (Jul. 2023).

\bibitem{castellani_spin-glass_2005}
T.~Castellani, A.~Cavagna, Spin-glass theory for pedestrians, Journal of Statistical Mechanics: Theory and Experiment 2005~(05) (2005) P05012.
\newblock \href {https://doi.org/10.1088/1742-5468/2005/05/P05012} {\path{doi:10.1088/1742-5468/2005/05/P05012}}.

\bibitem{bottcher_computational_2021}
L.~B{\"o}ttcher, H.~J. Herrmann, Computational {{Statistical Physics}}, Cambridge University Press, Cambridge, 2021.
\newblock \href {https://doi.org/10.1017/9781108882316} {\path{doi:10.1017/9781108882316}}.

\bibitem{corberi_kinetics_2023}
F.~Corberi, C.~Castellano, Kinetics of the one-dimensional voter model with long-range interactions (Sep. 2023).
\newblock \href {http://arxiv.org/abs/2309.16517} {\path{arXiv:2309.16517}}, \href {https://doi.org/10.48550/arXiv.2309.16517} {\path{doi:10.48550/arXiv.2309.16517}}.

\bibitem{helfmann_modelling_2023}
L.~Helfmann, N.~Djurdjevac~Conrad, P.~{Lorenz-Spreen}, C.~Sch{\"u}tte, Modelling opinion dynamics under the impact of influencer and media strategies, Scientific Reports 13~(1) (2023) 19375.
\newblock \href {https://doi.org/10.1038/s41598-023-46187-9} {\path{doi:10.1038/s41598-023-46187-9}}.

\bibitem{meyer_time_2024}
P.~G. Meyer, R.~Metzler, Time scales in the dynamics of political opinions and the voter model, New Journal of Physics 26~(2) (2024) 023040.
\newblock \href {https://doi.org/10.1088/1367-2630/ad27bc} {\path{doi:10.1088/1367-2630/ad27bc}}.

\end{thebibliography}





\end{document}


\begin{frontmatter}  
 
\title{ {\Large \textbf{Supplementary information} }  \\ 
Modeling public opinion control by a charismatic leader } 

\author[inst1]{Tiago S. A. N. Sim\~{o}es} 

\affiliation[inst1]{organization={University of Campania “Luigi Vanvitelli”, Department of Mathematics and Physics},
            addressline={Caserta, Viale Lincoln, 5, 81100},  
            country={Italy}}   

\author[inst2]{Antonio Coniglio} 
\affiliation[inst2]{organization={Università degli Studi di Napoli ‘‘Federico II’’, Physics Department},
            addressline={I-80126 Napoli}, 
            country={Italy}} 

\author[inst3,inst4]{Hans J. Herrmann}
             
\affiliation[inst3]{organization={Universidade Federal do Ceará, Departamento de Física},
            addressline={Fortaleza, Ceará, 60451-970},  
            country={Brazil}}   
\affiliation[inst4]{organization={ESPCI, PMMH},
            addressline={Paris, 7 quai St. Bernard, 75005 },   
            country={France}}     
 
\author[inst1]{Lucilla de Arcangelis}     
       
\end{frontmatter}

\beginsupplement       

\begin{figure}[h]            
    \centering
    \includegraphics[width=\linewidth]{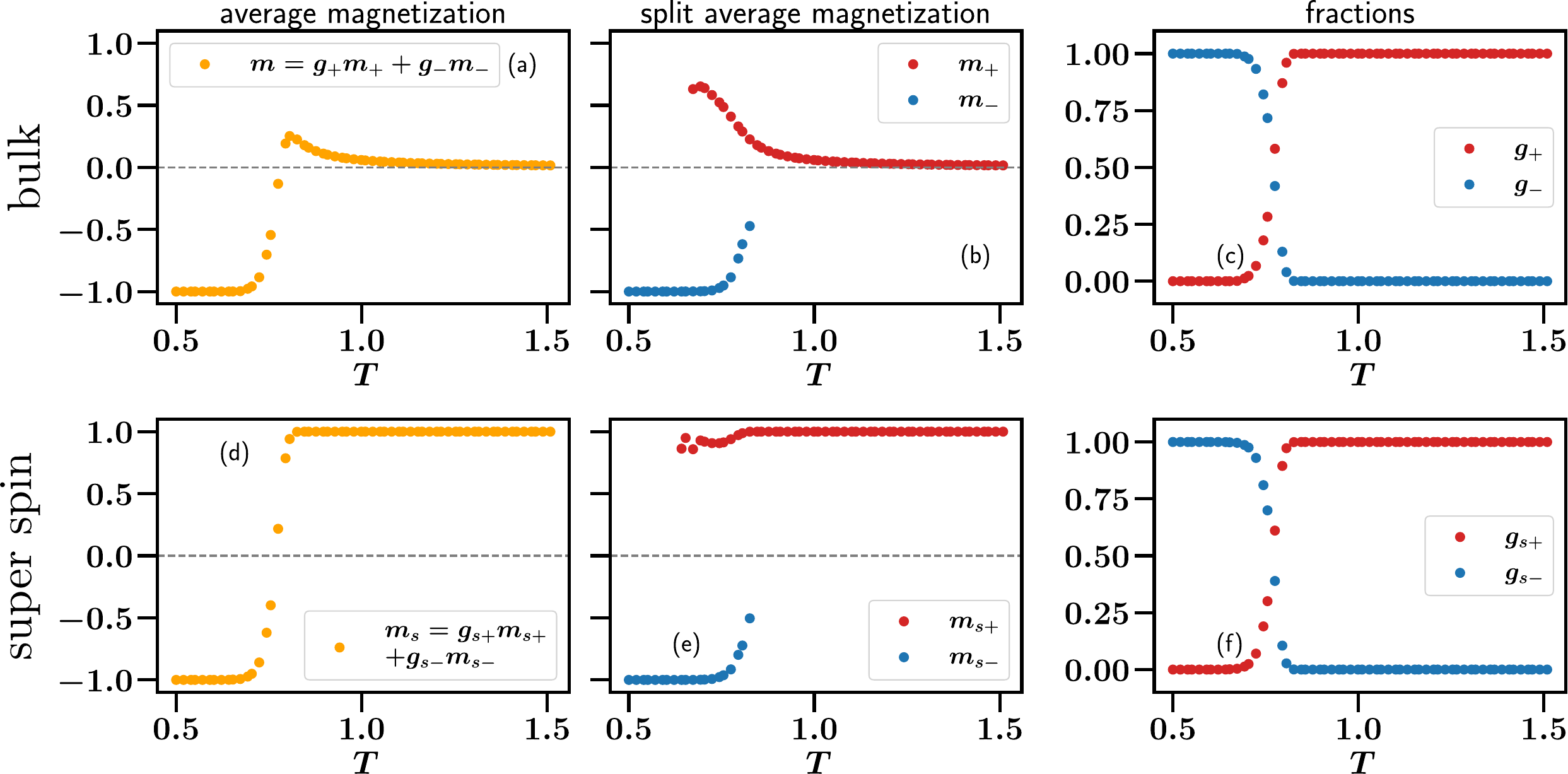}      
    \caption{\textbf{Results for the nearest neighbors generalized Ising model.} Results from Monte Carlo simulations of the average bulk $m$ and super spin $m_s$ magnetization (a and d), their partition into positive-negative average magnetization (b and e) and the fractions of Monte Carlo initial configurations which evolve towards a state with positive or negative average magnetization (c and f), as a function of $T \in \left[0.50,1.50\right]$, for the Ising model on a cubic lattice of side $L=10$ ($N \sim 10^3$) with periodic boundary conditions and nearest-neighbor interactions in the bulk instead of being fully-connected. For each temperature, results are averages over $ \NIC = 10^4 $ independent Monte Carlo simulations. The horizontal grey dashed lines indicate $m=0$ (a and b) and $m_s=0$ (d and e). Here we set the bulk interaction constant to $J=1/6$, since, besides the super spin, each bulk spin interacts only with its six nearest neighbors. The rest of the parameters are: $h = -1$, $\hsrd = 1 $, $J_s = 1.01 $.} 
    \label{fig_s:nn-Ising-1000}    
\end{figure}    

\begin{figure}[h]            
    \centering
    \includegraphics[width=\linewidth]{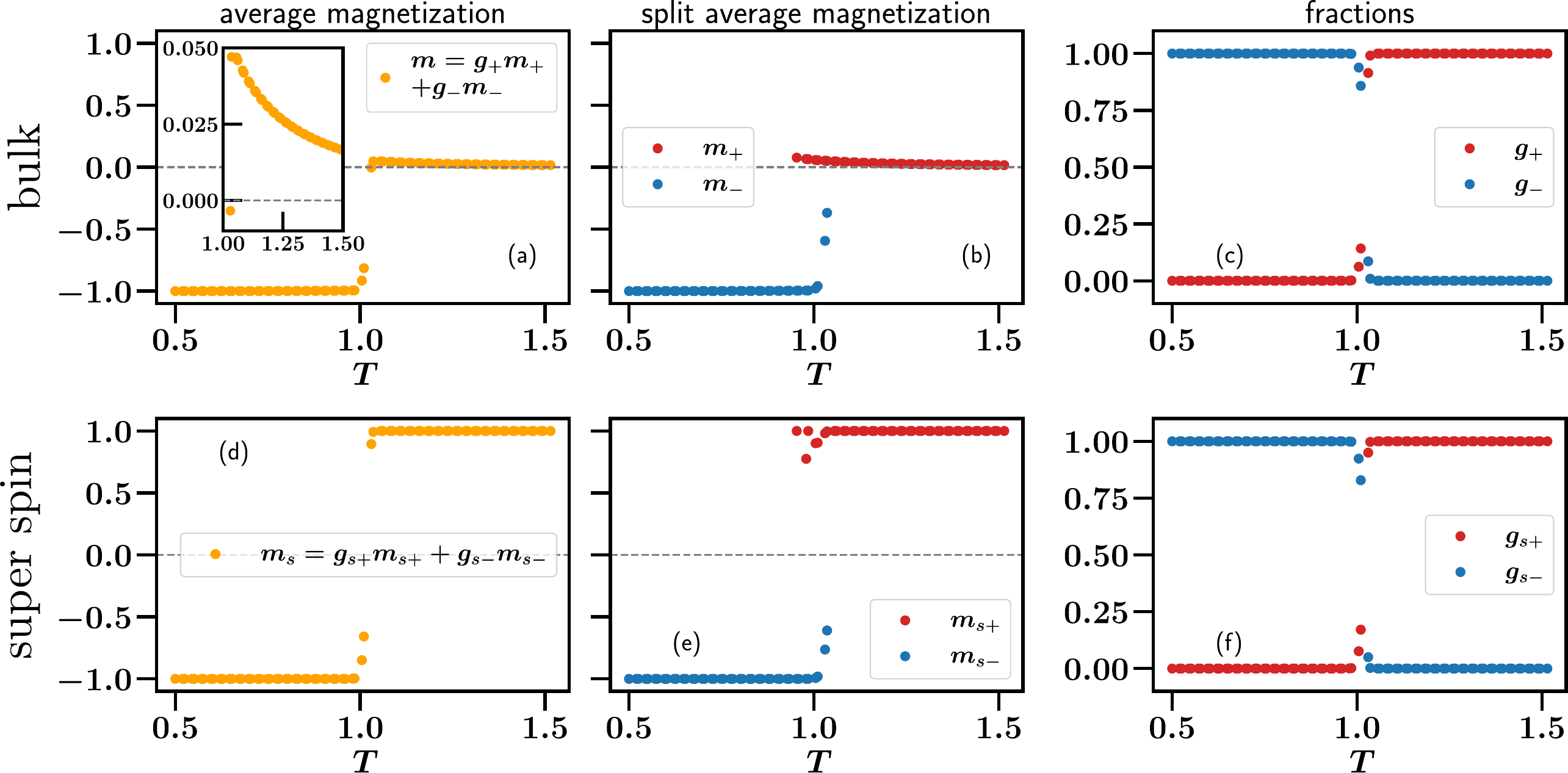}      
    \caption{\textbf{Results for the nearest neighbors generalized Ising model.} Same as in Fig. (\ref{fig_s:nn-Ising-10000}) but for a larger system size $L=22$ ($N \sim 10^4$). The inset in (a) is a zoom-in of the corresponding main plot near $T=1$, where the sign of $m$ changes.}   
    \label{fig_s:nn-Ising-10000}   
\end{figure}